\documentclass[pra,aps,groupedaddress,nofootinbib,12pt]{revtex4}
\usepackage{graphicx}

\usepackage{graphicx,amsfonts,amssymb,amsmath,hyperref,hypcap,enumerate}
\usepackage{slashed}
\usepackage{amsmath,amssymb,bm,graphicx,dsfont}
\usepackage[latin9]{inputenc}
\usepackage{amsmath}
\usepackage{slashed}
\usepackage{dsfont}
\usepackage{amstext}
\usepackage{amssymb}
\usepackage{amsbsy}
\usepackage{amsthm}
\usepackage{epsfig}
\usepackage{graphicx}
\usepackage{bbm}
\usepackage{hyperref}
\usepackage{color}
\usepackage{slashed}
\renewcommand{\v}[1]{\mathbf{#1}} 

\newcommand{\be}{\begin{equation}}
\newcommand{\ba}{\begin{align}}
\newcommand{\ee}{\end{equation}}
\newcommand{\bea}{\begin{eqnarray}}
\newcommand{\eea}{\end{eqnarray}}
\newcommand{\beq}{\begin{equation}}
\newcommand{\eeq}{\end{equation}}
\newcommand{\beqn}{\begin{eqnarray}}
\newcommand{\eeqn}{\end{eqnarray}}

\renewcommand{\vec}[1]{{\bf #1}}
\renewcommand{\hat}[1]{{\widehat #1}}

\begin{document}

\title{ Deconfined quantum critical points: a review}
	 
	\author{T. Senthil}\email{senthil@mit.edu}
	\affiliation{Department of Physics, Massachusetts Institute of Technology, Cambridge, MA, 02139}

\begin{abstract}
 Continuous phase transitions in equilibrium statistical mechanics were successfully described 50 years ago with the development of the 
renormalization group framework. This framework was initially developed in the context of phase transitions whose universal properties are captured by the long wavelength (and long time) fluctuations of a Landau order parameter field. Subsequent developments include a straightforward  generalization to a class of $T = 0$ phase transitions driven by quantum fluctuations. In the last 2 decades it has become clear that there is a vast landscape of quantum phase transitions 
where the physics is not always usefully (or sometimes cannot be) formulated in terms of fluctuations of a Landau order parameter field. A wide class of such 
phase transitions - dubbed deconfined quantum critical points - involve the emergence of fractionalized degrees of freedom coupled to 
emergent gauge fields. Here I review some salient aspects of these deconfined critical points. 
\end{abstract}


\date{\today}

\maketitle

\begin{center}
    To be published in ``50 years of the renormalization group", dedicated to the memory of Michael E. Fisher, edited by Amnon Aharony, Ora Entin-Wohlman, David Huse, and Leo Radzihovsky, World Scientific.
    \end{center}


\newpage

\section{Introduction: the Landau-Ginzburg-Wilson-Fisher paradigm}
The development of the theory of phase transitions in critical phenomena in the 1960s and 1970s is one of the towering achievements in modern physics. Michael Fisher played a leading role in this development. The resulting understanding, associated primarily with the names of Landau, Ginzburg, Wilson, and Fisher, has come to be known as the LGWF theory (or sometimes the LGWF paradigm). This theory - which we briefly review below - provided a conceptual framework to discuss equilibrium phase transition phenomena. Combined with renormalization group methods and the dimensionality expansion introduced by Wilson and Fisher, the theory enables computations of critical exponents and other universal properties in a systematic approximation. The LGWF theory was generalized in many different directions, eg, to deal with dynamical critical phenomena, and more pertinently to this chapter, to quantum phase transitions at $T = 0$. 

Phase transition theory begins by first asking how to distinguish distinct phases of matter.  A crucial idea (due to Landau) is to focus on the symmetries of the microscopic Hamiltonian describing the many body system. The equilibrium state in a range of parameters (or the ground state if, at $T = 0$) may not have the symmetries of the microscopic Hamiltonian, {\em i.e} the symmetry is spontaneously broken. Then this state is sharply distinct from a symmetry preserving state that may exist at other parameter values. The evolution from the broken symmetry phase to one where the symmetry is restored must then involve (at least one) phase transition. Landau also introduced the notion of an order parameter to quantify the extent of broken symmetry. The Landau order parameter appears as a new thermodynamic variable needed to fully characterize the macroscopic state of the symmetry broken phase. The universal properties of the phase transition are then associated  with the long wavelength (and long time, if dealing with quantum or dynamical critical phenomena) fluctuations of the order parameter field. The energy (or action, for quantum critical points) is written as a spatial integral of sums of an energy density that is expanded in powers of the order parameter field and its gradients. This is a coarse grained continuum description (an `effective field theory') of the original microscopic model. The statistical mechanics of the resulting action (the Ginzburg-Landau action) then provides a description of the universal physics of the critical point. In practice, this can be analysed within the framework of the renormalization group with the Wilson-Fisher $\epsilon$ expansion (or other expansions) as a tool to access the fixed points. 

These beautiful set of ideas are  sufficient to describe phase transitions in a large number of systems. The classic examples are the classical Ising or $O(N)$ magnets. However, in the last few decades, it has become clear that they turn out to not be the full story in a number of situations, particularly at quantum critical points. In this chapter we will survey a few different examples of how the LGWF paradigm breaks down. Nevertheless the framework of effective field theory and the renormalization group will remain central. What is different at these `beyond LGWF' transitions is that the effective field theory will not be the naive one written in terms of just the order parameter field. 

As we will see, there are a number of distinct routes for `beyond LGWF' phase transitions. First, there are phases of matter which cannot be simply characterized in terms of the usual broken symmetry paradigm and hence do not admit a description in terms of a Landau order parameter. Obviously, their phase transitions will not be described by the LGWF paradigm. More surprisingly, we can have Landau-forbidden quantum critical points between phases that themselves are described as broken symmetry phases with a Landau order parameter. Suppose the microscopic symmetry of some system is described by a group $G$, and consider two phases of matter where the symmetry $G$ is broken to a subgroup $H$ in one phase, and to a different subgroup $H'$ in another phase. In standard Landau theory. a second order phase transition between the two phases is possible only if one of the two groups $H, H'$ is a subgroup of the other. Remarkably, we have learnt that this rule can be broken at a variety of quantum crtical points\cite{deccp,deccplong}. The resulting critical theory is most usefully described in terms of a continuum field theory with emergent gauge fields coupled to matter fields carrying fractional quantum numbers of the microscopic global symmetry. These emergent fractionalized fields and associated gauge fields are not simply associated with quasiparticle excitations of either phase of matter; rather they rear their head at the critical point as useful variables to access the critical fixed point. Thus they have been dubbed `deconfined quantum critical points'.  

There are a number of examples of deconfined quantum critical points and related phenomena that have been studied in recent years.  For instance, examples have been found where the same phase transition allows multiple universality classes\cite{bi2019adventure}. These include situations\cite{bi2020landau} in which a Landau-allowed phase transition has both the conventional universality class as well as an unconventional one that is not described by the LGWF theory. Thus  even when the phase transition occurs between a trivial gapped phase and a broken symmetry one described by a Landau order parameter, the effective field theory of the critical point may not be captured by the LGWF action.   Even more striking are  ``unnecessary" critical points\cite{bi2019adventure} which do not separate distinct phases of matter but rather live suspended within a single phase of matter. It is possible to find a path in parameter space which avoids the critical point completely.  

Together these results underscore the idea that we have very little intuition for whether a phase transition is allowed to be second order, and if so, what its field theoretic description will be. As such they embolden us to contemplate the possibility that many mysterious quantum critical phenomena observed in metallic systems might find explanations in a form that is beyond a conventional description (known as the Hertz-Millis paradigm) in terms of electronic quasiparticles coupled to a critically fluctuating bosonic order parameter field.

\subsection{Phases of matter beyond Landau} 
For orientation, let us consider the different kinds of non-Landau phases of matter that are known to exist. The focus of this article will be on $T = 0$ quantum phases of matter, {\em i.e}, ground states of quantum many body systems.  A familiar example of phases whose distinction is not captured by the concept of broken symmetry  are  electronic band insulators  as compared to band metals.  The discovery of the integer and fractional quantum hall phases in the 1980s - which clearly do not break any microscopic symmetries - emphasized the possibility of new kinds of quantum ground states  which fall under the rubric of `topological quantum order'\footnote{In recent years, it has become popular to say that some such phases can actually be captured by the Landau paradigm if one generalizes the notion of symmetry. For a clear presentation of this point of view, see Ref. \cite{mcgreevy2023generalized}. For instance topologically ordered states of matter are considered to spontaneously break `higher-form' symmetries rather than ordinary ($0$-form) symmetries.  This point of view is interesting and useful. In applying it to the systems typically of interest in condensed matter physics, we must recognize that these higher form (or other generalized) symmetries are not present in the microscopic system. They are emergent symmetries in particular phases of matter that may then be spontaneously broken. This is different from the conventional Landau paradigm which dictates that we distinguish phases by asking whether {\em microscopic} symmetries are present/absent in the equilibrium (ground) state. Thus, in my view, incorporating topological ordered or other ``non-Landau" phases into the Landau paradigm requires not only generalizing the notion of symmetry but also generalizing what is meant by the Landau paradigm itself. Thus, in this article, I will refer to such phases as non-Landau as has been common for years.}. Since then, a large variety of `beyond-Landau' phases have been shown to be possible theoretically, and some have been discovered in experiment. These include other topologically ordered states (such as gapped quantum spin liquids in insulating magnets), symmetry protected topological states (such as topological band insulators and the Haldane spin-$1$ chain), and various gapless phases of matter where the gaplessness is not due to Goldstone fluctuations associated with a broken symmetry. The most famous example is the Landau fermi liquid itself. Other variants include Dirac materials such as graphene. Most novel are phases of matter with no quasiparticle excitations at all: examples include some gapless quantum spin liquids, and metallic non-fermi liquid phases. 

\begin{figure}
    \centering
    \includegraphics[width=\columnwidth]{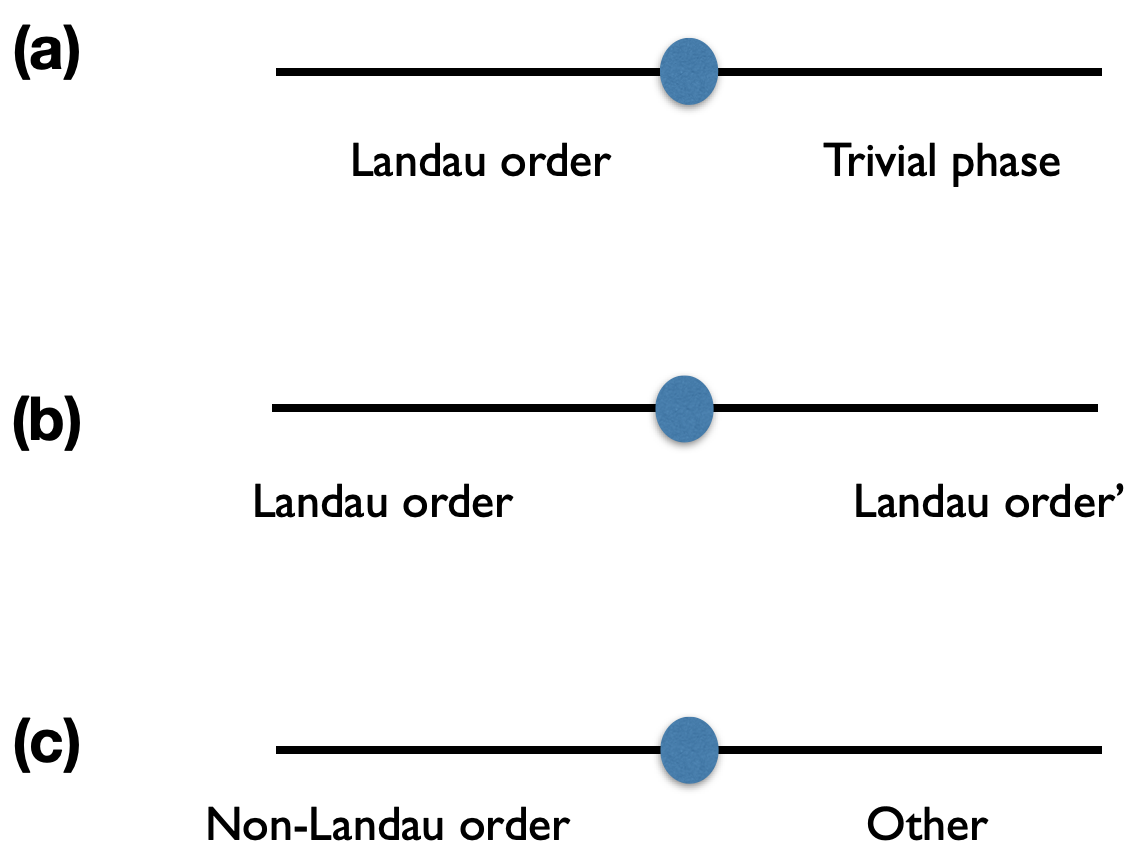}
    \caption{Varieties of quantum critical points. (a) The only case where the traditional LGWF paradigm is usually applicable. (b) Unless 
    the unbroken symmetry group of one phase is a subgroup of the unbroken symmetry of the other phase, the transition is Landau-forbidden (c) Non-Landau 
    order, if present in at least one of the two phases guarantees a transition beyond the LGWF paradigm.  }
    \label{fig:qcplscp}
\end{figure}

Obviously, if the phase itself is not described simply by specifying a Landau order parameter, the phase transition out of it will not be described by an LGWF action in terms of any fluctuating order parameter field. In Fig  \ref{fig:qcplscp} we sketch the variety of quantum phase transitions that accomodate this variety of phases of matter. The LGWF paradigm in its original form is only applicable to the situation where the symmetry preserving phase is gapped and trivial, and the symmetry broken phase is trivial apart from the broken symmetry. In all other cases, we should expect to find other descriptions of the phase transition. 

\subsection{Landau ordering transitions: a brief review}  

 We begin with a quick review of Landau ordering quantum phase transitions.  The canonical example is the transverse field Ising model on a $d$-dimensional lattice witha  global $Z_2$ symmetry. This has a disordered phase that preserves the global $Z_2$ symmetry, and an ordered phase that spontaneously breaks the symmetry. The quantum phase transition between these two phases is second order, and is in the universality class of the $d+1$-dimensional classical Ising model. The universal critical singularities are captured by an LGWF action of the form 
 \begin{equation}
 \label{lgwfaction} 
 {\cal S} = \int d^dx d\tau\left(\frac{\partial \phi}{\partial \tau}\right)^2 + \left(\mathbf{\nabla} \phi\right)^2 + r\phi^2 + u \phi^4 
 \end{equation} 
 where $\phi$ is the coarse-grained Ising order parameter field; $\tau$ is the imaginary-time coordinate, and $\mathbf{x}$ is the spatial coordinate. 

A simple generalization is to models with a global $O(N)$ symmetry which (for $d > 1$) have a second order phase transition between a symmetry preserving trivial gapped phase, and a symmetry broken phase where the global $O(N)$ symmetry is broken to the subgroup $O(N - 1)$. The transition is described by an action of the same form as in eqn. \ref{lgwfaction} but with $\phi$ replaced by an $N$-component order parameter field $\vec N$ that transforms as a vector under the global $O(N)$ symmetry. 

When might we expect the LGWF continuum field theory to correctly describe the phase transition out of the broken symmetry phase? Note that the LGWF theory should really be understood as an expansion about the disordered symmetry preserving phase which is implicitly assumed to be trivial and gapped. There are quasiparticle excitations charged under the global $O(N)$ symmetry. The most elementary such excitations transform under the vector representation of $O(N)$. As the transition is approached, the gap for these excitations closes. 

Thus we should expect LGWF actions of the form of Eqn. \ref{lgwfaction} to fail to describe the transition if the phase where the symmetry is restored is non-trivial in some way. For instance it may not be a trivial gapped phase and may have topological or other quantum order. Alternately it may break some other symmetry spontaneously. Clearly such cases are beyond the purview of the standard LGWF theory. 

An important and interesting class of examples arises in  situations where a trivial gapped symmetry preserving  phase is forbidden on general grounds. A classic example is in quantum many body systems of interacting particles with conserved particle number (corresponding to a global $U(1)$ symmetry) on a translation invariant lattice. If the lattice filling $\nu$ is not an integer, then a trivial symmetric gapped ground state is not possible through a (generalization\cite{oshikawa2000commensurability,hastings2004lieb} of) theorem by Lieb, Schultz, and Mattis (LSM). Such LSM restrictions apply very generally also to a number of translation-invariant models on a lattice where the degrees of freedom at each unit cell transform projectively under the global internal symmetry. For instance, in a spin system with global $SO(3)$ symmetry, if we have a spin-$1/2$ (or other half-integer spin) moment at each unit cell, then a trivial gapped ground state is forbidden by the LSM restriction. 

It follows that phase transitions in systems obeying LSM restrictions are never to be expected to be described by the LGWF theory. 

A closely related situation arises for phase transitions that occur at the boundary of a higher dimensional Symmetry Protected Topological\cite{senthil2015symmetry} (SPT) phase\footnote{As emphasized in Ref. \cite{cheng2016translational} we can regard a system satisying the LSM restriction itself as the boundary of a specific case of an STP phase in one higher dimension, one involving lattice translation as well as internal symmetries.}. An SPT phase has a ground state that cannot be adiabatically deformed through any path that preserves the global symmetries into a completely trivial ground state. Examples of SPT phases include topological band insulators in diverse dimensions and the spin-$1$ Haldane spin chain. In the last decade a great deal has been understood about the classification and physical properties of such SPT phases. A defining feature  is that it is not possible for the boundary of such an SPT phase to be trivially gapped. Thus phase transitions at SPT boundaries are also not expected to be described by the LGWF theory. Indeed the connection\cite{ashvinsenthil} of  the theory of deconfined quantum criticality to SPT boundaries has been very effective in understanding the latter. 

\section{LGWF* phase transitions} 
We begin by describing a simple example of a non-Landau transition, one that involves a critical point between a symmetry broken phase and one where the symmetry is restored but there is topological quantum order. Consider models of interacting bosons on a lattice with spatial dimension $d = 2$. The total boson number is conserved and there is a corresponding global $U(1)$ symmetry. At every lattice filling, a superfluid phase where the global $U(1)$ symmetry is spontaneously broken is clearly possible. As is well known, if the average number of bosons per unit cell is an integer, a  Mott insulating phase is also possible  where both the global $U(1)$ and translation symmetry are preserved. A physical picture is that in the Mott insulator, there are an integer number of bosons localized to each unit cell of the lattice. This Mott insulator is a symmetric trivial gapped phase. The corresponding superfluid-Mott phase transition is described by LGWF theory, and is in the universality class of the classical $XY$ model in $3$ dimensions. 

Next consider the situation when the number of bosons per unit cell is  $1/2$. Then a trivially gapped Mott insulator is forbidden. Nevertheless a symmetry preserving gapped ground state can exist so long as it has topological order and associated fractionally charged excitations. This state may be dubbed a fractionalized Mott insulator. The simplest example is when  the topological order is that of a deconfined $Z_2$ gauge theory. These can also be thought of as a quantum spin liquid ground state of a spin-$1/2$ magnet; for reviews, see Ref. \cite{savary2016quantum,broholm2020quantum} . This theory has gapped quasiparticles in distinct topological superselection sectors denoted $1, e, m$ and $\epsilon$. The corresponding $e$ and $m$ quasiparticles have bosonic self-statistics, and a mutual braiding statistics of $\pi$ ({\em i.e}, when one of $e,m$ is taken on a loop around the other, there is a phase of $\pi$). They may be identified with the  $Z_2$ gauge charge and  the $Z_2$ gauge flux of the $Z_2$ gauge theory\footnote{Their bound state $\epsilon$ is a fermion, and  has $\pi$ braiding statistics with both $e$ and with $m$; the superselection sector $1$ is simply associated with excitations that can be created by the action of local operators}. 
Though the $e$ and $m$ are both good quasiparticles in this state of matter, their non-local statistical interaction implies that neither can be created locally, {\em i.e} by acting with local operators in the physical Hilbert space. This allows them to carry fractional quantum numbers of the global symmetry. Though a single $e$ or $m$ particle is not a local excitation, $e^2$ or $m^2$ (where a pair of $e$ or $m$ particles are created) are local. Thus we can create, say, a pair of $e$ particles locally. The non-trivial statement is that the individual $e$-particles can be moved apart with a finite energy cost. Thus $e^2, m^2$ cannot carry fractional quantum numbers of the microscopic symmetries. It follows that $e$ can have  a quantum number $1/2$ under the global $U(1)$ symmetry. (In the presence of time reversal,  it can be shown\cite{ashvinsenthil,WS2013,MKF2013} that the $m$ particle cannot also have fractional $U(1)$ quantum numbers but will fractionalize lattice symmetries\cite{jalabert1991spontaneous,senthil2000z}.) Indeed it is possible to construct concrete models where precisely such $1/2$-charged $e$ particles, and neutral $m$ particles exist\cite{balents2002fractionalization,senthil2002microscopic,motrunich2002exotic}. Now if the microscopic boson model is at a lattice filling of $1/2$ a boson per unit cell, we can contemplate a fractionalized Mott insulator where there is a single $e$ particle that is localized in each unit cell. 

The phase transition between the superfluid and this fractionalized Mott insulator is conveniently described as a condensation of the $e$ particles while the gap to $m$ stays non-zero. At energy scales well below the $m$ gap, we can describe this transition through the action 
 \begin{equation}
 \label{lgwfaction*} 
 {\cal S} = \int d^dx d\tau\left|\frac{\partial \phi}{\partial \tau}\right|^2 + \left|\mathbf{\nabla} \phi\right|^2 + r|\phi|^2 + u |\phi|^4 
 \end{equation} 
where $\phi$ is a coarse-grained complex field whose quanta (in the disordered phase) describe the $e$ particles. This looks just the same as the standard LGWF action for the superfluid-Mott transition at integer filling but there is an important subtlety. As the $\phi$ field  creates the $e$-particle, it should not be directly identified with the superfluid order parameter, or indeed with any local observable in the original boson model. Relatedly this action does not know about the gapped $m$-sector which is part of the universal physics of both the fractionalized Mott insulator and the critical point. These subtleties can be handled by explicitly coupling the $\phi$ field to a $Z_2$ gauge field whose flux represents the $m$-particles. In that description, the $\phi$ field is not gauge invariant under $Z_2$ gauge transformations (though $\phi^2$ is). 

The action in Eqn. \ref{lgwfaction*} is nevertheless useful to read off many universal critical properties. We simply note that it describes the gapless sector as the $3D XY$ universality class except that operators which are not $Z_2$ gauge invariant are thrown out. The correlation length exponent $\nu$ (which is determined by the scaling dimension of $|\phi|^2$ which is gauge invariant) is the same as in the classical $3D$ $XY$ model. However the scaling dimension of the physical superfluid order parameter $\psi$ is different. This is because we can identify $\psi = \phi^2$. It thus has the anomalous exponent (introduced by Michael Fisher)  $\eta_\Psi \approx 1.49$. This large anomalous exponent must be contrasted with that of the usual 
superfluid-Mott transition at integer filling where $\eta \approx 0.03$. For numerical studies of this transition, see Ref. \cite{isakov2012universal}. 

Thus the the superfluid-fractionalized Mott insulator transition is beyond the standard LGWF paradigm, albeit with a simple relation to the standard LGWF theory. For these reasons, this kind of transition has come to be known as the LGWF* theory. 
For a sample of discussions of other LGWF* transitions, see Ref. \cite{chubukov1994universal,wang2021fractionalized,zhang2023xy}. 

Theories with a different structure also arise at certain phase transitions between topologically ordered and symmetry broken phases, for instance between fractional quantum Hall and superfluid phases of bosons on a periodic lattice. Unlike in the example above, there is usually no anyon in the fractional quantum Hall state that can be directly condensed to both break the global $U(1)$ symmetry and simultaneously destroy the topological order. Nevertheless, as argued in Ref. \cite{barkeshli2014continuous}, a continuous transition is possible and can be formulated in terms of emergent fermionic degrees of freedom coupled to a $U(1)$ gauge field. These fermions can be roughly thought of as composite fermions used to great effect in describing quantum Hall phenomena.

\section{Landau-forbidden quantum critical points between Landau allowed phases} 
We now turn to phase transitions between two phases which can both be characterized in terms of Landau order parameters corresponding to distinct broken symmetries. 
We will show that, contrary to naive expectations, such a phase transition can be second order. The primary example - which we review below - involves phase transitions in two dimensional insulating square lattice quantum antiferromagnets. These are described by models of systems of interacting quantum spins on the lattice.  A number of other examples in two space dimension have also been presented. (We postpone the discussion of other dimensions to later sections). An interesting example involves a phase transition in a system of interacting fermions, as we also review below. 

\subsection{Neel-VBS transition on the square lattice} 

Consider a system of $SU(2)$ spins on the square lattice with short range interactions. More precisely we will take the spin symmetry to be $SO(3) = \frac{SU(2)}{Z_2}$. and place spins that transform in the spin-$1/2$ (projective) representation of $SO(3)$ at each site.  We consider situations where the microscopic Hamiltonian has global $SU(2)$, time reversal, and square lattice space group symmetries.  

\begin{figure}
    \centering
    \includegraphics[width=\columnwidth]{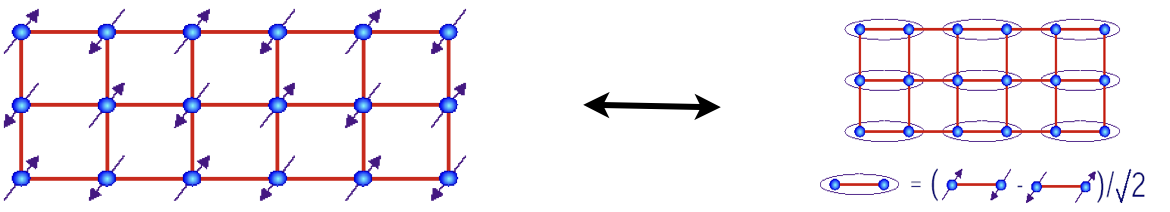}
    \caption{ The Neel and VBS states of a square lattice spin-$1/2$ magnet.}
    \label{fig:NeelVBS}
\end{figure}

The standard ground state of such a magnet is the N\'eel state  that breaks spin $SO(3)$  and some lattice symmetries. Specifically it preserves lattice rotation symmetry, and a combination of unit lattice translation and time reversal.  A different symmetry breaking state is a Valence Bond Solid (VBS) that preserves spin $SO(3)$, and time reversal but breaks unit lattice translations in one direction, and lattice rotational symmetry. Clearly these phases break distinct symmetries, and the remnant symmetry in neither phase is a subgroup of the other. Naively we might expect the transition between these two phases to either be first order, or go through a region where both broken symmetries coexist, or go through an intermediate phase with no broken symmetry. Note that LSM restrictions applied to these spin models tell us that the last possibility is only allowed if such an intermediate symmetric phase has topological or some other non-trivial quantum order. In particular a trivially gapped symmetric phase is not possible. Thus, as discussed above, a theory for the destruction of either the N\'eel or the VBS order will necessarily be `beyond LGWF'. 

In Ref. \cite{deccp,deccplong} it was argued that a direct second order transition is potentially possible and is described by the  field theory
\begin{equation}
\label{nccp1su2} {\cal L}_0 = \sum_{\alpha = 1,2} |D_b z_\alpha|^2
+ \left(|z_1|^2 + |z_2|^2\right)^2.
\end{equation}
Here $z_\alpha$ ($\alpha = 1,2$) are scalar fields (known as `bosonic spinons') that are   coupled to a
dynamical $U(1)$ gauge field $b$, and $D_{b,\mu}=\partial_{\mu}-ib_{\mu}$ is the covariant derivative. (This Euclidean action and  subsequent
similar actions are short-hands for the appropriate strongly
coupled  Wilson-Fisher critical theory where a background gauge
field has been promoted to a dynamical field.)   The model
has a global $SO(3)$ symmetry under which $z_\alpha$ transforms
as a spinor.\footnote{{Though $z_\alpha$ transforms as a spinor
under $SU(2)$, rotations in the center of $SU(2)$ can be
compensated by a $U(1)$ gauge transformation so that the spin
rotation symmetry of the model is $\frac{SU(2)}{Z_2} = SO(3)$. }}
 In the microsopic lattice spin model, this corresponds to the
$SO(3)$ spin rotation. It also has a global $U(1)$ symmetry
associated with the conservation of the flux  of $b$. In the microsopic lattice spin
model, this is not an exact symmetry. Consequently monopole
operators (which pick up a phase under a $U(1)$ rotation)
 must be added to the Lagrangian.
However, it is known that lattice symmetries ensure that the
minimal  allowed monopole operator (with continuum angular momentum $\ell = 0$) has
strength 4.

Analytic arguments \cite{deccp,deccplong} and numerical
calculations \cite{SandvikJQ,lousandvikkawashima,DCPscalingviolations} strongly support the
possibility that these monopoles are irrelevant at the putative critical
fixed point of Eq.~\eqref{nccp1su2}. The N\'eel phase is obtained
when $z_\alpha$ is condensed, and the VBS phase when $z_\alpha$ is
gapped. The N\'eel phase breaks $SO(3)$ to a $U(1)$ subgroup while
the VBS phase breaks the $U(1)$ flux conservation symmetry. The
N\'eel order parameter is simply $\v{N} = z^\dagger \v{\sigma} z$
($\v{\sigma}$ are Pauli matrices), and the VBS order parameter is
the strength 1 monopole operator ${\cal M}_b$ which creates $2\pi$
flux of $b$.

The field theory in Eqn. \ref{nccp1su2} in the absence of monopoles is known as the Non-Compact $CP^1$ ($NCCP^1$) model. It describes\cite{motrunich2004emergent} the phase transition in $O(3)$ models where hedgehog topological defects have been suppressed by hand. These hedgehog defects correspond precisely to monopoles in the $CP^1$ formulation of the $O(3)$ non-linear sigma model. 

A heroic body of numerical work on specific quantum magnets and related systems
\cite{SandvikJQ,melkokaulfan,lousandvikkawashima,Banerjeeetal,Sandviklogs,Kawashimadeconfinedcriticality,Jiangetal,deconfinedcriticalityflowJQ,DCPscalingviolations,emergentso5,MotrunichVishwanath2,kuklovetalDCPSU(2),Bartosch,CharrierAletPujol, Chenetal,Aletextendeddimer,powellmonopole} shows a
striking apparently continuous transition\footnote{It is however not yet clear whether the transition is truly second order or whether it displays only `quasiuniversal' behavior up to a very large but finite length scale \cite{Jiangetal, deconfinedcriticalityflowJQ,
Kawashimadeconfinedcriticality, Sandviklogs, kuklovetalDCPSU(2),sandvik2parameter,DCPscalingviolations,SimmonsDuffinSO(5),Nakayama,wang2021phases}. We will discuss this below.} between the N'eel and VBS phases with properties consistent with expectations based on the field theory in Eqn. \ref{nccp1su2}.  

What makes such a second order transition possible? The key physics is the properties of topological defects of the order parameters of either symmetry broken phase. To see this most simply, consider starting in the VBS phase which breaks the discrete symmetries of lattice translation and lattice rotation. There are four degenerate VBS ground states which are related to each other by $Z_4$ lattice rotations about a site.  Naively we might then expect a phase transition associated with the disordering of the VBS order to be in the universality class of the phase transition of a $Z_4$ clock model in $2+1-D$. As is well known, this is the same as the $2+1- D$ $XY$ universality class: four-fold anisotropy of the $XY$ order parameter is {\em dangerously irrelevant}\footnote{Another concept first introduced by Michael Fisher. } at the  $3D$ $XY$ fixed point. 

This naive expectation is in fact not correct due to non-trivial structure of the topological defects of the VBS order parameter. Before exploring this further, it is useful to first recall the physics of topological defects at the usual $2+1- D$ $XY$ phase transition. For an $XY$ order parameter, the topological defects are of course vortices. The transition out of the $XY$ ordered phase requires the proliferation of these vortices. The most famous example is in the BKT transition of the {\em classical} $2d$ $XY$ model but it remains true for the quantum $2+1-D$ $XY$ transition as well. Note that the $2+1-D$ $XY$ transition is described by the Wilson-Fisher fixed point with an action of the form in Eqn. \ref{lgwfaction*}. Equivalently we can describe it with a `dual' field theory in terms of a complex scalar field $\phi_v$  (identified physically with the vortex field) coupled to a dynamical $U(1)$ gauge field: 
\begin{equation}
 {\cal L}_{v} =  |D_b \phi_v|^2
+ \left(|\phi_v|^2  \right)^2.   
\end{equation}
This the celebrated particle-vortex duality\cite{peskin1978mandelstam,bosonvortexdh,bosonvortexfl}. The disordered phase of the $XY$ model is identified with the phase in which $\phi_v$ is condensed. This gaps out the gauge field $b$ through the Anderson-Higgs mechanism and we get a trivial gapped phase. The ordered phase of the $XY$ model is identified with the phase in which $\phi_v$ is gapped.  At low energies, we integrate out $\phi_v$ and get the free Maxwell action for the gauge field $b$. The corresponding `photon' is then identified with the Goldstone mode of the broken $XY$ symmetry. The $XY$ order parameter itself is identified with the monopole operator ${\cal M}_b$.  If the microscopic system has four-fold anisotropy on the $XY$ order parameter, then we are allowed to add the operator ${\cal M}_b^4 + c.c$ to the dual action. 

Armed with this understanding, let us now return to the task of disordering the VBS order in the square lattice quantum magnet. For more detail, please see  Ref. \cite{mlts04}.
Naturally the topological defects are domain walls. Four such elementary domain walls can end at a vortex which is a point defect.  It is easy to see that these VBS vortices necessarily involve a site of the lattice that is not paired with any other sites through a valence bond. Thus this site carries an unpaired spin-$1/2$ moment.  This is a key difference with an ordinary $Z_4$ clock model. Now consider quantum disordering the VBS order by proliferating ({\em i.e condensing}) these vortices. The spin-$1/2$ moment carried by these vortices will then lead to the breaking of spin rotation symmetry. It can be argued that the resulting state is precisely the N\'eel state. Indeed, following the principles of particle-vortex duality, the action in Eqn. \ref{nccp1su2} can be very simply understood as an  effective field theory for the VBS vortices which are identified with the $z_\alpha$ field\footnote{From a modern point of view, the structure of the topological defects implies that the $NCCP^1$ theory has a mixed ' t Hooft anomaly between the global $U(1)$ and $SO(3)$ symmetries\cite{wang2017deconfined}. Specifically, if we introduce a background gauge field $A$ for the global $U(1)$ symmetry, the monopole operators in $A$ will transform in the spinor representation of $SO(3)$ which is a reflection of the anomaly.   Similarly, the hypothesized emergent  $SO(5)$ symmetry discussed below of the low energy theory has its own 't Hooft anomaly which reduces to the mixed $U(1)-SO(3)$ anomaly upon restricting to that subgroup\cite{wang2017deconfined}. The anomaly perspective provides a powerful tool to obtain some non-perturbative constraints on the RG flows of the theory.  For a discussion on the relationship between the anomalies of the low energy theory and LSM constraints of the lattice spin model, see Ref. \cite{metlitski2018intrinsic,komargodski2018walls,ye2022topological}.  }. 

An alternate perspective is obtained by starting from the N\'eel state. 
The N\'eel order parameter $\vec N$ transforms in the vector representation of $SO(3)$. It thus admits skyrmion defects corresponding to $\Pi_2(S^2) = Z$. In the original lattice system, the conservation of skyrmions can potentially be violated by `hedgehog' events in space-time.  However for spin-$1/2$ magnets on the square lattice, it can be shown that single skyrmions transform non-trivially under lattice symmetries\cite{HaldaneBerry,ReSaSUN,deccp,deccplong}. Thus a single skyrmion cannot be created or destroyed while preserving microscopic symmetries. In fact, the symmetry transformation of the skyrmion creation operator is identical to that of the VBS order parameter. If we now disorder the N\'eel order by proliferating skyrmions, the resulting paramagnetic phase will have VBS order.   

The structure of the topological defects also reveals a problem with a naive Ginzburg-Landau action to describe the competition between the N\'eel and VBS orders. Indeed, a description of this competition will focus on the N\'eel order parameter $\vec N$ and the VBS order parameter $\psi$, and write an action for their fluctuations. To that end it is convenient to introduce a $5$-component unit vector $n^a$ ($a = 1,...., 5$)
such that $n^{3,4,5}$ correspond to the three components of the
N\'eel vector $\vec N$, and $n^{1,2}$ to the two real components of the VBS
order parameter $\psi$. A naive Ginzburg-Landau action will then take the form of a non-linear sigma model 
\begin{equation}
    {\cal S}_{naive} =  \frac{1}{2g}  \int d^3x \, (\partial  n^a)^2  + \lambda \left(\vec N^2 - |\psi|^2\right) + .....
\end{equation}
The first term is fully $O(5)$ symmetric under  rotations of $n^a$; the second term introduces anisotropy between the $5$ components of $n^a$ so that the symmetry is reduced to $O(3) \times O(2)$. This term can also be used to tune between N\'eel and VBS ordered states. The ellipses include terms that further break these symmetries to those of the lattice spin model.  This kind of Ginzburg-Landau action would usually suggest that the N\'eel-VBS transition is either first order, or goes through an intermediate phase with neither order. Further in this naive sigma model, this intermediate phase will be a trivially gapped phase\footnote{A direct second-order Neel-VBS transition can then only be expected at a multicritical point.} However since such a phase is forbidden  in the lattice spin-$1/2$ magnet, clearly the naive sigma model needs to be modified in some way. 

The crucial point is that the naive Ginzburg-Landau model does not capture the structure of the topological defects described above. Consider for instance the VBS ordered state where $n^a$ points in the $n^4, n^5$ plane, and construct a vortex configuration of $\psi = n^4 + i n^5$. At the core of the vortex, $|\psi| = 0$ and so the unit vector $n^a$ will point along the N\'eel direction. The dynamics of the N\'eel vector in the vortex core will be described by an $SO(3)$ rotor model and will have quantum states transforming in integer representations of $SO(3)$. However we saw above that the VBS vortex in our physical setting must transform as spin-$1/2$ under $SO(3)$. This leads to a `quantum intertwinement' of the two orders that must be incorporated into the Ginzburg-Landau theory. The correct action\cite{tanakahu,tsmpaf06} turns out to require inclusion of a Wess-Zumino-Witten (WZW) term at level-$1$ for the dynamics of the $SO(5)$ vector $n^a$: 
 \begin{equation}
S =  \frac{1}{2g}  \int d^3x \, (\partial  n^a)^2 + 2\pi
\Gamma\left[ n^a \right].
\end{equation}
The WZW term $\Gamma$ is defined in the standard way: the field
$n^a$ defines a map from spacetime $S^3$ to the target space
$S^4$, and $\Gamma$ is the ratio of the volume in $S^4$ traced out
by $n_a$ to the total volume of $S^4$. If $n^a(x,u)$ is any smooth
extension of $n^a(x)$ such that $n^a(x,0) = (0,0,0,0,1)$ and
$n^a(x,1) = n^a(x)$, then
\begin{equation}
\Gamma  =  \frac{\epsilon_{abcde}}{\textrm{Area}(S^4)}
\int_{0}^{1} d\,u \int d^3 x \, n^a\partial_x n^b \partial_y n^c
\partial_t n^d \partial_u n^e.
\end{equation}
The WZW term correctly captures
the non-trivial quantum numbers of the topological defects and is
responsible for the non-Landau physics. Repeating the calculation of the structure of the VBS vortex, it is easy to see that the WZW term reduces - at the vortex core - to the standard spin Berry phase as appropriate for a spin-$1/2$ degree of freedom deescribed by $\vec N$.   

 The sigma model is well-defined as a continuum field theory only in the weak coupling limit, where it is
ordered. Here there is a clear semiclassical picture for the
effect of the WZW term on the
topological defects in the ordered state. For the transition itself --- driven by  the $\lambda$ term  --- this ordered state corresponds to a first
order phase transition.  In particular the sigma model formulation does not by itself make it at all natural that the N\'eel-VBS transition can be second order. To directly say anything about the putative second-order Neel-VBS phase transition in this formulation requires extending the sigma model to strong coupling, and looking for a symmetry preserving critical phase (see below for discussion of a recent novel numerical study\cite{zhou2023mathrm}). At strong coupling the sigma model theory is
non-renormalizable and requires an alternative formulation as a
continuum field theory. Physically, disordered phases of the sigma
model (defined with an explicit UV cutoff) correspond to phases
where topological defects of the order parameter have
proliferated. Thus a modification of the topological defects leads
to a modification of the corresponding disordered phases. The sigma
model formulation thus exposes the seed, in the ordered phase, of
the impending non-Landau physics of the disordered critical
regime.

The field theory in Eqn.\ref{nccp1su2} thus raises the possibility that the Neel-VBS transition may be second order. Whether it is actually so or not depends on the IR fate of the $NCCP^1$ model.  This is however hard to study.  To get a handle, it is extremely useful to consider the general case of interacting $SU(N)$ quantum spins.  The proposed field theory for the non-Landau transition can be studied analytically in the limit of large-$N$ through a $1/N$ expansion. A second order transition described by a conformal field theory is obtained. Thus the lattice spin model for sufficiently large-$N$ provides a solvable and, hence important, example of a non-Landau deconfined quantum critical point. Extrapolating to small values of $N$, the critical fixed point has large values of the anomalous exponent $\eta$ for both the N'eel and VBS order parameters. The  four-fold monopoles  are irrelevant at the critical fixed point, but of course they are important in the VBS ordered phase where they pin the orientation of the VBS order parameter to one of four degenerate values. This dangerous irrelevance of four-fold monopoles leads to the presence of two divergent length scales: one is the spin correlation length $\xi$, and the other - which may be identified as a scale associated with the pinning of the VBS order - $\xi_{VBS}$ diverges as a power of $\xi$.

The field theory for the $N = 2$ model can only be studied numerically, and its precise fate remains murky. In particular, many studies\cite{Jiangetal, deconfinedcriticalityflowJQ,
Kawashimadeconfinedcriticality, Sandviklogs, kuklovetalDCPSU(2),sandvik2parameter,DCPscalingviolations,SimmonsDuffinSO(5),Nakayama,zhao2022scaling,da2023teaching} notice vexing problems (such as drifts with increasing system size in some critical properties). 

Further studies\cite{emergentso5} show the
emergence of an $SO(5)$ symmetry that rotates the N\'eel and VBS
order parameters into one another. Such a symmetry is of course suggested by the sigma model formulation of the intertwinement betweeen Neel and VBS orders, which however is not directly suited to describe the QCP as discussed above. 
This symmetry is however not at all apparent in the $NCCP^1$ formulation. Ref. \cite{wang2017deconfined} provided an understanding of the emergence of the $SO(5)$ symmetry from many points of view.  The field theory admits multiple (`dual') descriptions. In any one of these descriptions only a subgroup of the full symmetry is apparent but it is a different subgroup in different descriptions. Thus by combining the information from these multiple descriptions we can get an understanding of the full emergent $SO(5)$ symmetry.


A manifestly $SO(5)$ invariant effective field theory\cite{wang2017deconfined} of the intertwinement of the N\'eel and VBS orders is  obtained by starting with a fermionic spinon description of the square lattice spin-$1/2$ magnet. This naturally leads to a low energy theory of two flavors of massless Dirac fermions coupled to a dynamical $SU(2)$ gauge field --- a theory denoted $N_f = 2$ $QCD_3$. It is expected\footnote{It is a bosonic theory with global $SO(5)$ symmetry, and the same anomaly.} that this theory  shares the same universal infrared physics as the $NCCP^1$ theory. 

The possibility that the putative fixed point theory may have full $SO(5)$ symmetry enables another check on its existence as a conformal field theory using 
conformal bootstrap methods. Unfortunately an $SO(5)$ symmetric CFT with exponents in the range expected from numerical calculations of  lattice spin models is not found\cite{Nakayama}. Combined with the drifts in critical exponents mentioned above, it appears possible that for $N = 2$, the Neel-VBS transition is weakly first order but with a very long correlation length leading to the appearance of a continuous transition 

A possible explanation of these facts (an unnaturally weak first order transition together with the emergence of the $SO(5)$ symmetry) is provided by the idea that the transition in the $N = 2$ model is `pseudocritical'\cite{nahum2015deconfined,wang2017deconfined}. Pseudocriticality refers to a situation where the transition is second order `close' to the physical model (in a generalized space including spatial dimensionality $d$, number of components $N$, etc) but the fixed points move slightly away from the real axis for the physical model. Then along the renormalization group flow the coupling constants linger (`walk') for a long time when they are closest to the fixed point but never really reach it, and ultimately flow away. Pseudocriticality commonly occurs when a stable fixed point merges with an unstable one, and they annihilate each other. A famous example\cite{nienhuispotts}, is the 5-state classical Potts model in two dimensions. For other examples and a general discussion, see Ref. \cite{gorbenko2018walking}. For the N\'eel-VBS transition, the pseudocritical scenario finds support\cite{ma2020theory,nahum2020note} in an epsilon expansion near space dimension $d = 1$ of the WZW non-linear sigma model. 

An alternate suggestion\cite{zhao2020multicritical,liu2022emergence,lu2021self,yang2022quantum,shackleton2021deconfined} is that at $N = 2$, there is (in a parameter range) a quantum spin liquid intervening between the N'eel and VBS phases. As the parameters are varied, the quantum spin liquid region shrinks and is eventually replaced by a first order N'eel-VBS transition. There is an associated multicritical point when the spin liquid terminates and the line of first order transitions begins, and this is identified\cite{yang2022quantum} with the deconfined quantum critical point. 

Very recently, a beautiful study\cite{zhou2023mathrm} provides strong support for the emergence of approximate conformal invariance and pseudocriticality at this phase transition. This work studied the  $SO(5)$ WZW model regulated as a fermion model on a ``fuzzy sphere" (a sphere with a magnetic monopole at the center, which leads to the formation of Landau levels for the fermions). As shown in a previous work\cite{zhu2023uncovering} on the $2+1-D$ Ising transition , such a fuzzy sphere regularization enables examining the emergence of conformal symmetry by studying the spectrum of eigenstates through exact diagonalization. For conformal field theories, these energies are related to operator scaling dimensions, and the presence of conformal symmetry leads to many restrictions on the spectrum. For the regulated $SO(5)$ WZW model, the fuzzy sphere calculation finds a spectrum nicely consistent with the emergence of conformal symmetry. However with increasing system size, an $SO(5)$ singlet operator becomes weakly relevant. As we emphasized above, for the N\'eel-VBS transition, it is necessary that the $SO(5)$ WZW model at strong coupling have a critical phase (the phase transition is then tuned by an anisotropy term that explicitly breaks $SO(5)$ to $SO(3) \times O(2)$).  The presence of a weakly relevant $SO(5)$ singlet is thus nicely consistent with the conjectured pseudocriticality at the N\'eel-VBS transition. 

Clearly further clarifying the precise fate of the $N = 2$ model is an important goal for the near future. Nevertheless we emphasize that the existence of a second order N\'eel-VBS transition for large enough $N$ establishes the matter-of-principle question on whether such transitions can exist at all.

\subsection{Other realizations} 
A different realization of a deconfined critical point is a phase transition in a system of interacting electrons. The continuous global internal symmetry is $U(2)$ corresponding to conservation of the total electronic charge and spin. In additon there is time reversal symmetry under which the electrons transform as Kramers doublets. Consider a phase  where $SU(2)$ spin rotation symmetry is broken spontaneously to a $U(1)$ subgroup but the charge $U(1)$ and time reversal are preserved.   The corresponding order parameter is a three component field that transforms in the spin-$1$ representation of the $SU(2)$ spin rotation symmetry. Now let us further specialize to a situation where - at the mean field level - there is a quantum spin Hall effect\cite{kane2005a,kane2005b}. In the bulk of such a  system, the electronic spectrum is gapped but there will be gapless edge states at a boundary to the vacuum. 

It was shown in Ref. \cite{grover2008topological} that fat skyrmion defects of the order parameter carry electrical charge-$2$. Restoring the broken $SO(3)$ symmetry by condensing skyrmion defects then leads to superconducting order\footnote{This `skyrmion condensation' mechanism of superconductivity has, in recent years, been proposed to be realized in twisted bilayer graphene\cite{khalaf2021charged,christos2020superconductivity}. It has been demonstrated numerically both on lattice models\cite{liu2019superconductivity}, and on continuum models similar to those appropriate for twisted bilayer graphene\cite{chatterjee2022skyrmion}.}.  {\em i.e}, to a spontaneous breakdown of the global $U(1)$ associated with charge conservation. The phase transition between these two states is described at low energies by the $NCCP^1$ field theory, and hence is another realization of the deconfined quantum critical point.  In the $NCCP^1$ formulation, the flux of the $U(1)$ gauge field corresponds to the physical charge density, and hence is conserved exactly. Thus monopole operators of the $U(1)$ gauge field correspond to Cooper pair operators that change the charge by $2$. Obviously these monopole operators are not allowed in the action (unlike in the quantum magnetism realization)\footnote{The astute reader will recall that the low energy $NCCP^1$ field theory has a mixed t' Hooft anomaly between the global $U(1)$ and the global $SO(3)$ symmetries.   While we are certainly used to emergent symmetries having 't Hooft anomalies, it may be surprising that there is such an anomaly in a system where these symmetries are well-defined as `on-site' symmetries of a microscopic lattice model. The resolution is that the anomaly is only present if the theory is regarded as a bosonic theory.  In the  presence of the (gapped) electronic degrees of freedom, the anomaly is trivialized, and this allows an on-site realization of the $SO(3) \times U(1)$ symmetry. This has the consequence\cite{thorngren2021intrinsically,ma2022edge} that in a system with boundaries, the fermion will be gapless at the edge even in the presence of the gapless critical bulk.   }. 

The absence of monopoles of any strength implies that (in contrast to the Neel-VBS example), here there is only a single diverging length scale at the transition. Fermionic lattice models that realize this transition have been written down and are amenable to large scale Monte Carlo studies\cite{liu2019superconductivity,hohenadler2022thermodynamic} which find evidence for a second order transition.

Deconfined critical field theories also  arise in the context of phase transitions between trivial and SPT phases \cite{ashvinsenthil,kevinQSH,so4qsh}. These are often related by duality transformations to the theories of the Landau-forbidden 
transitions\cite{wang2017deconfined}. 
 
\section{DQCP in 1d} 
In the last few years, there have been a number of interesting studies\cite{jiang2019ising,roberts2019deconfined,huang2019emergent,zhang2023exactly,mudry2019quantum,su2023boundary} on one dimensional lattice quantum models that possess deconfined quantum critical points. In contrast to their $2$-dimensional counterparts, these one dimensional models can be studied reliably with field theoretic and numerical methods. Models have even been constructed which are exactly solvable and show a deconfined quantum critical point. 

Consider a $1d$ lattice model\cite{jiang2019ising} of spin-$1/2$ moments  with the Hamiltonian 
\begin{equation} 
{\cal H} = - \sum_j \left(J_x \sigma^x_j \sigma^x_{j+1} + J_z \sigma^z_j \sigma^z_{j+1} \right) + \sum_j \left(K_x \sigma^x_j \sigma^x_{j+2} + K_z \sigma^z_j \sigma^z_{j+2} \right)
\end{equation} 
The model has two global $Z_2$ symmetries corresponding to $\pi$ rotations of all the spins about either the $x$ or the $z$-axis. These are denoted $Z_2^z$ and $Z_2^x$ respectively. It also has lattice translation symmetry, and the (anti-unitary) time reversal symmetry. By tuning the various parameters $J_{x,z}, K_{x,z}$, a number of phases can be accessed.  For instance, a   $z-FM$ phase spontaneously breaks the $Z_2^x$  and time reversal symmetries but preserves $Z_2^z$ and translation. A distinct VBS phase preserves both $Z_2$ symmetries and time reversal but breaks unit lattice translation symmetry. The phase transition between these two phases can be shown to be second order. 
The structure  of the resulting theory has many similarities with that of the Neel-VBS transition in two space dimensions\cite{jiang2019ising}. The Landau-forbidden phase transition is enabled by the structure of the topological defects. In the VBS phase, a domain wall leaves behind an unpaired site with a single spin-$1/2$ moment. This unpaired spin transforms projectively under the $Z_2^x \times Z_2^y$ symmetry. Destroying the Neel order by proliferating these domain walls then necessarily breaks $Z_2^x \times Z_2^y$ symmetry.

Ref. \cite{zhang2023exactly} constructed an exactly solvable $1d$ model with global $Z_2 \times Z_2$ symmetry which features a continuous phase transition between two phases which break the two different $Z_2$ symmetries. The symmetry action is however anomalous (one of the $Z_2$ symmetries is not an on-site symmetry). Thus this system is usefully viewed as living at the boundary of a $2d$ Symmetry Protected Topological phase of spins with the same $Z_2 \times Z_2$ symmetry. The model could be exactly mapped to a $Z_4$ clock model coupled to a $Z_2$ gauge field. In $1d$ the presence of the $Z_2$ gauge field simply implies global selection rules on physical states when the system is placed on a ring. The Landau-forbidden transition can then be related to the ordering transition of the $Z_4$ model which is continuous.

\section{DQCP in 3d: solvable examples} 
\begin{figure}
    \centering
    \includegraphics[width=\columnwidth]{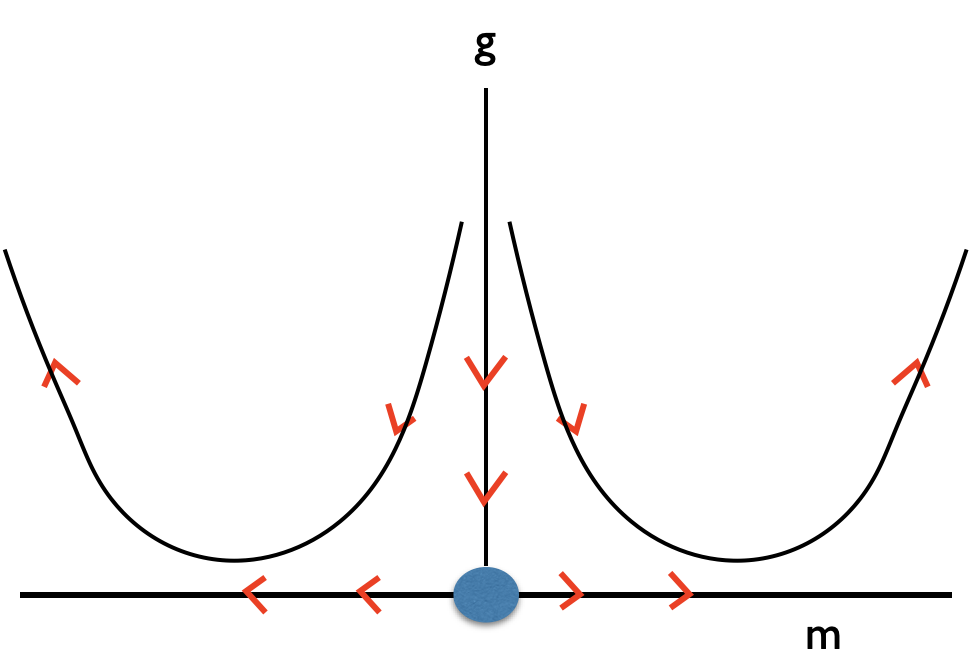}
    \caption{ Renormalization group flows of the IR-free non-abelian gauge theory.}
    \label{fig:3ddqcpRG}
\end{figure}

A fruitful approach to generating a wealth of examples of deconfined criitcal points whose universal IR physics is solvable is described in Ref. \cite{bi2019adventure}. The idea is to study deconfined critical theories that are free in the IR but become strongly coupled and confine upon moving away from criticality. To illustrate this, consider a quantum field theory in $3+1$ spacetime dimensions described by $N_f$ massless Dirac fermions coupled to, say, an $SU(2)$ gauge field. The one-loop beta function for the gauge coupling $g$ is well known, and takes the form
\begin{equation}
\beta(g^2)=\frac{dg^2}{dl}= \frac{g^3}{24\pi^2}  (N_f - 11)   
\end{equation} 
The theory is thus IR-free for $N_f > 11$. The gauge coupling $g$ is irrelevant at the IR fixed point.  The low energy physics is then given by a theory of free massless Dirac fermions (`quarks'), and a separate set of free bosons (`gluons'). Now consider perturbing the theory by adding a mass $m$ for the Dirac fermions. If we demand time reversal symmetry, then there is a unique (real) mass term that is allowed. However once the fermions are massive, they can be integrated out and the low energy theory is that of pure $SU(2)$ gauge theory; in this phase, $g$ flows to strong coupling, and the theory is confining. The RG flows are shown in Fig. \ref{fig:3ddqcpRG}. The massless point $m = 0$ can be interpreted as a quantum critical point in the phase diagram. The nature of the two phases depends on the value of $N_f$. For even $N_f$, the $m = 0$ point describes a phase transition between a trivial gapped phase and an SPT phase\footnote{This field theory should be regarded as describing a microscopic system of interacting bosons: all local, {\em i.e} gauge invariant, operators in the theory are bosons. See Ref. \cite{bi2019adventure} for identification of the protecting symmetries and further details.} (we discuss odd $N_f$ below).  Though this model may seem esoteric to condensed matter physics, it has the advantage that it realizes a particularly simple and solvable (in the IR) example of deconfined quantum criticality.

\section{Other novel QCPs} 
In this section, we briefly discuss work in the last few years demonstrating a number of interesting critical phenomena. These were first found in the context of IR-free deconfined critical theories of the kind described above (by considering different gauge groups and different matter content). More natural models (from a condensed matter perspective) that realize some of these phenomena have since been found. 
\subsection{Multiple universality classes} 
In traditional phase transition theory, it is common to assume that,  given two phases that are separated by a second order transition, there is a unique universality class for the phase transition. This is not generally true. For instance in disordered systems ({\em eg}, the $\pm J$ Ising spin glass), it has been known for a long time that there may be multiple universality classes for the same phase transition. It is perhaps to be expected that this phenomenon of multiple universality classes also occurs in clean systems but the present author had not seen examples till recently. The IR-free non-abelian gauge theories naturally lead to many examples of multiple (deconfined critical) universality classes for the same trivial-SPT phase transition. For a simpler example in a $d = 1$ spin systems, see Ref. \cite{prakash2022multiversality}. 

\subsection{Landau beyond Landau} 
The examples of multi-universality mentioned above were for transitions between gapped symmetry preserving phases. These of course are not to be described within the LGWF paradigm. It is thus interesting that it also possible to find examples where a Landau ordering phase transition that admits a universality class within the LGWF paradigm also allows a distinct universality class that is beyond LGWF. An example of this phenomenon is found in the model of the IR-free non-abelian gauge theory. Specifically,  the $SU(2)$ gauge theory with $N_f$ massless fermions, with $N_f$ odd and sufficiently large, is a (deconfined) critical point between a trivial gapped phase and a phase that spontaneously breaks time reversal symmetry but is otherwise gapped. The transition is driven by changing the sign of the fermion mass. This transition can clearly also be in the usual Ising universality class. The deconfined critical universality class has (many) more gapless degrees of freedom which further transform non-trivially under some global symmetries\footnote{These global symmetries are unbroken on either side of the phase transition; in the usual Ising universality class, excitations charged under these other symmetries will be gapped. The deconfined universality class however has gapless excitations that are charged under these other global symmetries.} - other than time reversal - present in this system. There is thus no formulation of this new universality class in terms of a theory written just in terms of the Ising order parameter field; it thus represents a dramatic breakdown of the LGWF paradigm. 

At the time this article was written, there are no other examples of this `Landau beyond Landau' phenomenon in simpler models in condensed matter physics, and finding one such is a fascinating future endeavor.

 \subsection{``Unnecessary"  critical points}
 It is often stated - in discussions on quantum materials - that the most fundamental question about a quantum critical point is to ask about what distinguishes the phases on either side. However, 
Ref. \cite{bi2019adventure} described an interesting phenomenon where a second order critical line lives ``inside" a single phase of matter. This was dubbed an ``unnecessary" quantum critical point.  In this case there is no sharp distinction between the two sides of the transition and it is possible to go around it smoothly. It is analogous to the liquid-gas transition except that it is second order. Furthermore in the initial example described in Ref. \cite{bi2019adventure}, this unnecessary critical point had emergent deconfined gapless degrees of freedom though the phase it is submerged in is a trivial gapped phase. This example was once again constructed by studying the phase diagram of perturbed IR-free non-abelian gauge theories. 

Subsequent work has found a number of examples in simpler systems. These include spin chains\cite{thorngren2021intrinsically,prakash2022multiversality} in $1d$, and free fermion systems\cite{bi2019adventure,jian2020generic} in $2d$ or $3d$. It appears that unnecessary criticality is not an uncommon phenomenon. 
 
\section{Discussion} 
As reviewed here, there are a number of phase transitions where the  strategy of focusing on long wavelength, long time fluctuations of a Landau order parameter field is not adequate. Nevertheless the framework of  effective field theory and the renormalization group remain crucial. This is arguably one of the most important insights of the theory of critical phenomena developed in the 1960s and 1970s. At the deconfined critical points described in this review, useful effective field theories can be constructed in terms of emergent gapless fields that are charged under an emergent gauge field. 

There are many further developments and applications that we have not discussed. For instance, recent work\cite{shu2023equilibration}  studies the dynamics of the Neel-VBS transition in a driven $2d$ quantum magnet through numerical simulations. The deconfinement at the putative critical point impacts the annealing process, and leads to dynamics distinct from that described by the usual Kibble-Zurek scaling of conventional phase transitions. 

An outgrowth of the theory of deconfined criticality is the understanding that there can be stable gapless `deconfined' phases in two dimensional quantum magnets whose low energy theories involve fermions coupled to gauge fields\cite{hermele2004stability}. An interesting example is a phase known as an  algebraic (or Dirac) spin liquid where there are emergent gapless fermions coupled to a dynamical $U(1)$ gauge field. Interesting generalizations of these states have been proposed\cite{zou2021stiefel} where it is not clear that there is {\em any} Lagrangian-based field theoretic description at all. If such states do exist, they may pose a challenge even to the paradigm of writing down an action for an effective field theory to describe the low energy physics. 

At the time of writing, there are relatively few experimental realizations of deconfined quantum critical points. For the Neel-VBS transition, very recent experiments on the material $SrCu_2(BO_3)_2$ have been interpreted in terms of a deconfined critical point\cite{cui2023proximate}. A provocative recent paper suggests\cite{song2023unconventional} that a deconfined critical point may underlie the onset of superconductivity in the monolayer material $WTe_2$. A realization of  a $1d$ deconfined quantum critical point in Rydberg atom simulators has been proposed\cite{lee2022landau}. 

One of the prime motivations for the original development of the theory of deconfined criticality in Ref. \cite{deccp,deccplong} was the observation of non-fermi liquid physics at the transition between two fermi liquid phases in the heavy fermion materials\cite{coleman2001topical,lohneysen2007fermi}. These critical points are usually  associated with the onset of magnetic order out of the heavy fermi liquid but there may be a concomitant destruction\cite{si2001locally,senthil2004weak} of the Kondo screening responsible for the formation of the heavy fermi liquid in the first place. There is of course a qualitative resemblance to the simpler deconfined critical points discussed above. However a theory of these kinds of  heavy fermion critical points must deal with hard questions on the fate of the fermi surfaces in either phase. Despite years of effort, these critical points remain mysterious. Nevertheless it is encouraging that the study of simpler models (without fermi surfaces) already demonstrates that Landau order parameter fluctuations may not be the full story at phase transitions, and, even worse, may distract from the essential physics of some phase transitions. 

This work was supported by NSF grant DMR-2206305, and partially through a Simons Investigator Award from the Simons Foundation. Part of this article was written at the KITP and  supported in part by the National Science Foundation under Grant No. NSF PHY-1748958. I am honored to be part of an academic lineage which places me as Michael Fisher's academic `great grandson' (through David Nelson and Subir Sachdev).  A different connection is through my postdoctoral advisor (Matthew Fisher) - one of Michael's physicist sons. Reading an article on critical phenomena and the renormalization group by Michael was one of the reasons I got attracted to condensed matter physics as an undergraduate. It is thus a pleasure to contribute to this memorial volume. 
 
 \bibliographystyle{ws-rv-van}
\bibliography{deccprev}

\begin{thebibliography}{102}
\providecommand{\natexlab}[1]{#1}
\providecommand{\url}[1]{\texttt{#1}}
\expandafter\ifx\csname urlstyle\endcsname\relax
  \providecommand{\doi}[1]{doi: #1}\else
  \providecommand{\doi}{doi: \begingroup \urlstyle{rm}\Url}\fi

\bibitem{deccp}
T.~Senthil, A.~Vishwanath, L.~Balents, S.~Sachdev, and M.~P.~A. Fisher,
  Deconfined quantum critical points, \emph{Science}. {\bf 303}, \penalty0
  1490,  (2004).

\bibitem{deccplong}
T.~Senthil, L.~Balents, S.~Sachdev, A.~Vishwanath, and M.~P.~A. Fisher, Quantum
  criticality beyond the landau-ginzburg-wilson paradigm, \emph{Phys. Rev. B}.
  {\bf 70}, \penalty0 144407 (Oct, 2004).
\newblock \doi{10.1103/PhysRevB.70.144407}.
\newblock URL \url{http://link.aps.org/doi/10.1103/PhysRevB.70.144407}.

\bibitem{bi2019adventure}
Z.~Bi and T.~Senthil, Adventure in topological phase transitions in 3+ 1-d:
  non-abelian deconfined quantum criticalities and a possible duality,
  \emph{Physical Review X}. {\bf 9}\penalty0 (2), \penalty0 021034,  (2019).

\bibitem{bi2020landau}
Z.~Bi, E.~Lake, and T.~Senthil, Landau ordering phase transitions beyond the
  landau paradigm, \emph{Physical Review Research}. {\bf 2}\penalty0 (2),
  \penalty0 023031,  (2020).

\bibitem{mcgreevy2023generalized}
J.~McGreevy, Generalized symmetries in condensed matter, \emph{Annual Review of
  Condensed Matter Physics}. {\bf 14}, \penalty0 57--82,  (2023).

\bibitem{oshikawa2000commensurability}
M.~Oshikawa, Commensurability, excitation gap, and topology in quantum
  many-particle systems on a periodic lattice, \emph{Physical review letters}.
  {\bf 84}\penalty0 (7), \penalty0 1535,  (2000).

\bibitem{hastings2004lieb}
M.~B. Hastings, Lieb-schultz-mattis in higher dimensions, \emph{Phys. Rev. B}.
  {\bf 69}, \penalty0 104431 (Mar, 2004).
\newblock \doi{10.1103/PhysRevB.69.104431}.
\newblock URL \url{https://link.aps.org/doi/10.1103/PhysRevB.69.104431}.

\bibitem{senthil2015symmetry}
T.~Senthil, Symmetry-protected topological phases of quantum matter,
  \emph{Annu. Rev. Condens. Matter Phys.} {\bf 6}\penalty0 (1), \penalty0
  299--324,  (2015).

\bibitem{cheng2016translational}
M.~Cheng, M.~Zaletel, M.~Barkeshli, A.~Vishwanath, and P.~Bonderson,
  Translational symmetry and microscopic constraints on symmetry-enriched
  topological phases: A view from the surface, \emph{Physical Review X}. {\bf
  6}\penalty0 (4), \penalty0 041068,  (2016).

\bibitem{ashvinsenthil}
A.~{Vishwanath} and T.~{Senthil}, {Physics of Three-Dimensional Bosonic
  Topological Insulators: Surface-Deconfined Criticality and Quantized
  Magnetoelectric Effect}, \emph{Physical Review X}. 3\penalty0 (1):\penalty0
  011016 (Jan., 2013).
\newblock \doi{10.1103/PhysRevX.3.011016}.

\bibitem{savary2016quantum}
L.~Savary and L.~Balents, Quantum spin liquids: a review, \emph{Reports on
  Progress in Physics}. {\bf 80}\penalty0 (1), \penalty0 016502,  (2016).

\bibitem{broholm2020quantum}
C.~Broholm, R.~Cava, S.~Kivelson, D.~Nocera, M.~Norman, and T.~Senthil, Quantum
  spin liquids, \emph{Science}. {\bf 367}\penalty0 (6475), \penalty0 eaay0668,
  (2020).

\bibitem{WS2013}
C.~Wang and T.~Senthil, Boson topological insulators: A window into highly
  entangled quantum phases, \emph{Phys. Rev. B}. {\bf 87}, \penalty0 235122
  (Jun, 2013).
\newblock \doi{10.1103/PhysRevB.87.235122}.
\newblock URL \url{http://link.aps.org/doi/10.1103/PhysRevB.87.235122}.

\bibitem{MKF2013}
M.~A. Metlitski, C.~L. Kane, and M.~P.~A. Fisher, Bosonic topological insulator
  in three dimensions and the statistical witten effect, \emph{Phys. Rev. B}.
  {\bf 88}, \penalty0 035131 (Jul, 2013).
\newblock \doi{10.1103/PhysRevB.88.035131}.
\newblock URL \url{http://link.aps.org/doi/10.1103/PhysRevB.88.035131}.

\bibitem{jalabert1991spontaneous}
R.~A. Jalabert and S.~Sachdev, Spontaneous alignment of frustrated bonds in an
  anisotropic, three-dimensional ising model, \emph{Physical Review B}. {\bf
  44}\penalty0 (2), \penalty0 686,  (1991).

\bibitem{senthil2000z}
T.~Senthil and M.~P. Fisher, Z 2 gauge theory of electron fractionalization in
  strongly correlated systems, \emph{Physical Review B}. {\bf 62}\penalty0
  (12), \penalty0 7850,  (2000).

\bibitem{balents2002fractionalization}
L.~Balents, M.~P. Fisher, and S.~M. Girvin, Fractionalization in an easy-axis
  kagome antiferromagnet, \emph{Physical Review B}. {\bf 65}\penalty0 (22),
  \penalty0 224412,  (2002).

\bibitem{senthil2002microscopic}
T.~Senthil and O.~Motrunich, Microscopic models for fractionalized phases in
  strongly correlated systems, \emph{Physical Review B}. {\bf 66}\penalty0
  (20), \penalty0 205104,  (2002).

\bibitem{motrunich2002exotic}
O.~Motrunich and T.~Senthil, Exotic order in simple models of bosonic systems,
  \emph{Physical review letters}. {\bf 89}\penalty0 (27), \penalty0 277004,
  (2002).

\bibitem{isakov2012universal}
S.~V. Isakov, R.~G. Melko, and M.~B. Hastings, Universal signatures of
  fractionalized quantum critical points, \emph{Science}. {\bf 335}\penalty0
  (6065), \penalty0 193--195,  (2012).

\bibitem{chubukov1994universal}
A.~V. Chubukov, T.~Senthil, and S.~Sachdev, Universal magnetic properties of
  frustrated quantum antiferromagnets in two dimensions, \emph{Physical review
  letters}. {\bf 72}\penalty0 (13), \penalty0 2089,  (1994).

\bibitem{wang2021fractionalized}
Y.-C. Wang, M.~Cheng, W.~Witczak-Krempa, and Z.~Y. Meng, Fractionalized
  conductivity and emergent self-duality near topological phase transitions,
  \emph{Nature Communications}. {\bf 12}\penalty0 (1), \penalty0 5347,  (2021).

\bibitem{zhang2023xy}
Y.-H. Zhang, Z.~Zhu, and A.~Vishwanath, Xy* transition and extraordinary
  boundary criticality from fractional exciton condensation in quantum hall
  bilayer, \emph{arXiv preprint arXiv:2302.03703}.  (2023).

\bibitem{barkeshli2014continuous}
M.~Barkeshli and J.~McGreevy, Continuous transition between fractional quantum
  hall and superfluid states, \emph{Physical Review B}. {\bf 89}\penalty0 (23),
  \penalty0 235116,  (2014).

\bibitem{SandvikJQ}
A.~W. Sandvik, Evidence for deconfined quantum criticality in a two-dimensional
  heisenberg model with four-spin interactions, \emph{Phys. Rev. Lett.} {\bf
  98}, \penalty0 227202 (Jun, 2007).
\newblock \doi{10.1103/PhysRevLett.98.227202}.
\newblock URL \url{http://link.aps.org/doi/10.1103/PhysRevLett.98.227202}.

\bibitem{lousandvikkawashima}
J.~Lou, A.~W. Sandvik, and N.~Kawashima, Antiferromagnetic to
  valence-bond-solid transitions in two-dimensional $\text{SU}(n)$ heisenberg
  models with multispin interactions, \emph{Phys. Rev. B}. {\bf 80}, \penalty0
  180414 (Nov, 2009).
\newblock \doi{10.1103/PhysRevB.80.180414}.
\newblock URL \url{http://link.aps.org/doi/10.1103/PhysRevB.80.180414}.

\bibitem{DCPscalingviolations}
A.~Nahum, J.~T. Chalker, P.~Serna, M.~Ortu\~no, and A.~M. Somoza, Deconfined
  quantum criticality, scaling violations, and classical loop models,
  \emph{Phys. Rev. X}. {\bf 5}, \penalty0 041048 (Dec, 2015).
\newblock \doi{10.1103/PhysRevX.5.041048}.
\newblock URL \url{http://link.aps.org/doi/10.1103/PhysRevX.5.041048}.

\bibitem{motrunich2004emergent}
O.~I. Motrunich and A.~Vishwanath, Emergent photons and transitions in the o
  (3) sigma model with hedgehog suppression, \emph{Physical Review B}. {\bf
  70}\penalty0 (7), \penalty0 075104,  (2004).

\bibitem{melkokaulfan}
R.~G. Melko and R.~K. Kaul, Scaling in the fan of an unconventional quantum
  critical point, \emph{Phys. Rev. Lett.} {\bf 100}, \penalty0 017203 (Jan,
  2008).
\newblock \doi{10.1103/PhysRevLett.100.017203}.
\newblock URL \url{http://link.aps.org/doi/10.1103/PhysRevLett.100.017203}.

\bibitem{Banerjeeetal}
A.~Banerjee, K.~Damle, and F.~Alet, Impurity spin texture at a deconfined
  quantum critical point, \emph{Phys. Rev. B}. {\bf 82}, \penalty0 155139 (Oct,
  2010).
\newblock \doi{10.1103/PhysRevB.82.155139}.
\newblock URL \url{http://link.aps.org/doi/10.1103/PhysRevB.82.155139}.

\bibitem{Sandviklogs}
A.~W. Sandvik, Continuous quantum phase transition between an antiferromagnet
  and a valence-bond solid in two dimensions: Evidence for logarithmic
  corrections to scaling, \emph{Phys. Rev. Lett.} {\bf 104}, \penalty0 177201
  (Apr, 2010).
\newblock \doi{10.1103/PhysRevLett.104.177201}.
\newblock URL \url{http://link.aps.org/doi/10.1103/PhysRevLett.104.177201}.

\bibitem{Kawashimadeconfinedcriticality}
K.~Harada, T.~Suzuki, T.~Okubo, H.~Matsuo, J.~Lou, H.~Watanabe, S.~Todo, and
  N.~Kawashima, Possibility of deconfined criticality in su($n$) heisenberg
  models at small $n$, \emph{Phys. Rev. B}. {\bf 88}, \penalty0 220408 (Dec,
  2013).
\newblock \doi{10.1103/PhysRevB.88.220408}.
\newblock URL \url{http://link.aps.org/doi/10.1103/PhysRevB.88.220408}.

\bibitem{Jiangetal}
F.-J. Jiang, M.~Nyfeler, S.~Chandrasekharan, and U.-J. Wiese, From an
  antiferromagnet to a valence bond solid: evidence for a first-order phase
  transition, \emph{Journal of Statistical Mechanics: Theory and Experiment}.
  {\bf 2008}\penalty0 (02), \penalty0 P02009,  (2008).
\newblock URL \url{http://stacks.iop.org/1742-5468/2008/i=02/a=P02009}.

\bibitem{deconfinedcriticalityflowJQ}
K.~Chen, Y.~Huang, Y.~Deng, A.~B. Kuklov, N.~V. Prokof'ev, and B.~V. Svistunov,
  Deconfined criticality flow in the heisenberg model with ring-exchange
  interactions, \emph{Phys. Rev. Lett.} {\bf 110}, \penalty0 185701 (May,
  2013).
\newblock \doi{10.1103/PhysRevLett.110.185701}.
\newblock URL \url{http://link.aps.org/doi/10.1103/PhysRevLett.110.185701}.

\bibitem{emergentso5}
A.~Nahum, P.~Serna, J.~T. Chalker, M.~Ortu\~no, and A.~M. Somoza, Emergent
  so(5) symmetry at the n\'eel to valence-bond-solid transition, \emph{Phys.
  Rev. Lett.} {\bf 115}, \penalty0 267203 (Dec, 2015).
\newblock \doi{10.1103/PhysRevLett.115.267203}.
\newblock URL \url{http://link.aps.org/doi/10.1103/PhysRevLett.115.267203}.

\bibitem{MotrunichVishwanath2}
O.~I. {Motrunich} and A.~{Vishwanath}, {Comparative study of Higgs transition
  in one-component and two-component lattice superconductor models},
  \emph{ArXiv e-prints} (May. 2008).

\bibitem{kuklovetalDCPSU(2)}
A.~B. Kuklov, M.~Matsumoto, N.~V. Prokof'ev, B.~V. Svistunov, and M.~Troyer,
  Deconfined criticality: Generic first-order transition in the su(2) symmetry
  case, \emph{Phys. Rev. Lett.} {\bf 101}, \penalty0 050405 (Aug, 2008).
\newblock \doi{10.1103/PhysRevLett.101.050405}.
\newblock URL \url{http://link.aps.org/doi/10.1103/PhysRevLett.101.050405}.

\bibitem{Bartosch}
L.~Bartosch, Corrections to scaling in the critical theory of deconfined
  criticality, \emph{Phys. Rev. B}. {\bf 88}, \penalty0 195140 (Nov, 2013).
\newblock \doi{10.1103/PhysRevB.88.195140}.
\newblock URL \url{http://link.aps.org/doi/10.1103/PhysRevB.88.195140}.

\bibitem{CharrierAletPujol}
D.~Charrier, F.~Alet, and P.~Pujol, Gauge theory picture of an ordering
  transition in a dimer model, \emph{Phys. Rev. Lett.} {\bf 101}, \penalty0
  167205 (Oct, 2008).
\newblock \doi{10.1103/PhysRevLett.101.167205}.
\newblock URL \url{http://link.aps.org/doi/10.1103/PhysRevLett.101.167205}.

\bibitem{Chenetal}
G.~Chen, J.~Gukelberger, S.~Trebst, F.~Alet, and L.~Balents, Coulomb gas
  transitions in three-dimensional classical dimer models, \emph{Phys. Rev. B}.
  {\bf 80}, \penalty0 045112 (Jul, 2009).
\newblock \doi{10.1103/PhysRevB.80.045112}.
\newblock URL \url{http://link.aps.org/doi/10.1103/PhysRevB.80.045112}.

\bibitem{Aletextendeddimer}
D.~Charrier and F.~Alet, Phase diagram of an extended classical dimer model,
  \emph{Phys. Rev. B}. {\bf 82}, \penalty0 014429 (Jul, 2010).
\newblock \doi{10.1103/PhysRevB.82.014429}.
\newblock URL \url{http://link.aps.org/doi/10.1103/PhysRevB.82.014429}.

\bibitem{powellmonopole}
G.~J. Sreejith and S.~Powell, Scaling dimensions of higher-charge monopoles at
  deconfined critical points, \emph{Phys. Rev. B}. {\bf 92}, \penalty0 184413
  (Nov, 2015).
\newblock \doi{10.1103/PhysRevB.92.184413}.
\newblock URL \url{http://link.aps.org/doi/10.1103/PhysRevB.92.184413}.

\bibitem{sandvik2parameter}
H.~Shao, W.~Guo, and A.~W. Sandvik, Quantum criticality with two length scales,
  \emph{Science}. {\bf 352}\penalty0 (6282), \penalty0 213--216,  (2016).
\newblock ISSN 0036-8075.
\newblock \doi{10.1126/science.aad5007}.
\newblock URL \url{http://science.sciencemag.org/content/352/6282/213}.

\bibitem{SimmonsDuffinSO(5)}
D.~Simmons-Duffin.
\newblock unpublished.

\bibitem{Nakayama}
Y.~Nakayama and T.~Ohtsuki, Necessary condition for emergent symmetry from the
  conformal bootstrap, \emph{Phys. Rev. Lett.} {\bf 117}, \penalty0 131601
  (Sep, 2016).
\newblock \doi{10.1103/PhysRevLett.117.131601}.
\newblock URL \url{http://link.aps.org/doi/10.1103/PhysRevLett.117.131601}.

\bibitem{wang2021phases}
Z.~Wang, M.~P. Zaletel, R.~S. Mong, and F.~F. Assaad, Phases of the (2+ 1)
  dimensional so (5) nonlinear sigma model with topological term,
  \emph{Physical Review Letters}. {\bf 126}\penalty0 (4), \penalty0 045701,
  (2021).

\bibitem{peskin1978mandelstam}
M.~E. Peskin, Mandelstam-'t hooft duality in abelian lattice models,
  \emph{Annals of Physics}. {\bf 113}\penalty0 (1), \penalty0 122--152,
  (1978).

\bibitem{bosonvortexdh}
C.~Dasgupta and B.~I. Halperin, Phase transition in a lattice model of
  superconductivity, \emph{Phys. Rev. Lett.} {\bf 47}, \penalty0 1556--1560
  (Nov, 1981).
\newblock \doi{10.1103/PhysRevLett.47.1556}.
\newblock URL \url{http://link.aps.org/doi/10.1103/PhysRevLett.47.1556}.

\bibitem{bosonvortexfl}
M.~P.~A. Fisher and D.~H. Lee, Correspondence between two-dimensional bosons
  and a bulk superconductor in a magnetic field, \emph{Phys. Rev. B}. {\bf 39},
  \penalty0 2756--2759 (Feb, 1989).
\newblock \doi{10.1103/PhysRevB.39.2756}.
\newblock URL \url{http://link.aps.org/doi/10.1103/PhysRevB.39.2756}.

\bibitem{mlts04}
M.~{Levin} and T.~{Senthil}, {Deconfined quantum criticality and N{\'e}el order
  via dimer disorder}, \emph{\prb}. 70\penalty0 (22):\penalty0 220403 (Dec.,
  2004).
\newblock \doi{10.1103/PhysRevB.70.220403}.

\bibitem{wang2017deconfined}
C.~Wang, A.~Nahum, M.~A. Metlitski, C.~Xu, and T.~Senthil, Deconfined quantum
  critical points: symmetries and dualities, \emph{Physical Review X}. {\bf
  7}\penalty0 (3), \penalty0 031051,  (2017).

\bibitem{metlitski2018intrinsic}
M.~A. Metlitski and R.~Thorngren, Intrinsic and emergent anomalies at
  deconfined critical points, \emph{Physical Review B}. {\bf 98}\penalty0 (8),
  \penalty0 085140,  (2018).

\bibitem{komargodski2018walls}
Z.~Komargodski, T.~Sulejmanpasic, and M.~{\"U}nsal, Walls, anomalies, and
  deconfinement in quantum antiferromagnets, \emph{Physical Review B}. {\bf
  97}\penalty0 (5), \penalty0 054418,  (2018).

\bibitem{ye2022topological}
W.~Ye, M.~Guo, Y.-C. He, C.~Wang, and L.~Zou, Topological characterization of
  lieb-schultz-mattis constraints and applications to symmetry-enriched quantum
  criticality, \emph{SciPost Physics}. {\bf 13}\penalty0 (3), \penalty0 066,
  (2022).

\bibitem{HaldaneBerry}
F.~D.~M. Haldane, O(3) nonlinear $\ensuremath{\sigma}$ model and the
  topological distinction between integer- and half-integer-spin
  antiferromagnets in two dimensions, \emph{Phys. Rev. Lett.} {\bf 61},
  \penalty0 1029--1032 (Aug, 1988).
\newblock \doi{10.1103/PhysRevLett.61.1029}.
\newblock URL \url{http://link.aps.org/doi/10.1103/PhysRevLett.61.1029}.

\bibitem{ReSaSUN}
N.~Read and S.~Sachdev, Spin-peierls, valence-bond solid, and n\'eel ground
  states of low-dimensional quantum antiferromagnets, \emph{Phys. Rev. B}. {\bf
  42}, \penalty0 4568--4589 (Sep, 1990).
\newblock \doi{10.1103/PhysRevB.42.4568}.
\newblock URL \url{http://link.aps.org/doi/10.1103/PhysRevB.42.4568}.

\bibitem{tanakahu}
A.~Tanaka and X.~Hu, Many-body spin berry phases emerging from the
  $\ensuremath{\pi}$-flux state: Competition between antiferromagnetism and the
  valence-bond-solid state, \emph{Phys. Rev. Lett.} {\bf 95}, \penalty0 036402
  (Jul, 2005).
\newblock \doi{10.1103/PhysRevLett.95.036402}.
\newblock URL \url{http://link.aps.org/doi/10.1103/PhysRevLett.95.036402}.

\bibitem{tsmpaf06}
T.~Senthil and M.~P.~A. Fisher, Competing orders, nonlinear sigma models, and
  topological terms in quantum magnets, \emph{Phys. Rev. B}. {\bf 74},
  \penalty0 064405 (Aug, 2006).
\newblock \doi{10.1103/PhysRevB.74.064405}.
\newblock URL \url{http://link.aps.org/doi/10.1103/PhysRevB.74.064405}.

\bibitem{zhou2023mathrm}
Z.~Zhou, L.~Hu, W.~Zhu, and Y.-C. He, The so(5) deconfined phase transition
  under the fuzzy sphere microscope: Approximate conformal symmetry,
  pseudo-criticality, and operator spectrum, \emph{arXiv preprint
  arXiv:2306.16435}.  (2023).

\bibitem{zhao2022scaling}
J.~Zhao, Y.-C. Wang, Z.~Yan, M.~Cheng, and Z.~Y. Meng, Scaling of entanglement
  entropy at deconfined quantum criticality, \emph{Physical Review Letters}.
  {\bf 128}\penalty0 (1), \penalty0 010601,  (2022).

\bibitem{da2023teaching}
Y.~Da~Liao, G.~Pan, W.~Jiang, Y.~Qi, and Z.~Y. Meng, The teaching from
  entanglement: 2d deconfined quantum critical points are not conformal,
  \emph{arXiv preprint arXiv:2302.11742}.  (2023).

\bibitem{nahum2015deconfined}
A.~Nahum, J.~Chalker, P.~Serna, M.~Ortu{\~n}o, and A.~Somoza, Deconfined
  quantum criticality, scaling violations, and classical loop models,
  \emph{Physical Review X}. {\bf 5}\penalty0 (4), \penalty0 041048,  (2015).

\bibitem{nienhuispotts}
B.~Nienhuis, A.~N. Berker, E.~K. Riedel, and M.~Schick, First- and second-order
  phase transitions in potts models: Renormalization-group solution,
  \emph{Phys. Rev. Lett.} {\bf 43}, \penalty0 737--740 (Sep, 1979).
\newblock \doi{10.1103/PhysRevLett.43.737}.
\newblock URL \url{http://link.aps.org/doi/10.1103/PhysRevLett.43.737}.

\bibitem{gorbenko2018walking}
V.~Gorbenko, S.~Rychkov, and B.~Zan, Walking, weak first-order transitions, and
  complex cfts, \emph{Journal of High Energy Physics}. {\bf 2018}\penalty0
  (10), \penalty0 1--49,  (2018).

\bibitem{ma2020theory}
R.~Ma and C.~Wang, Theory of deconfined pseudocriticality, \emph{Physical
  Review B}. {\bf 102}\penalty0 (2), \penalty0 020407,  (2020).

\bibitem{nahum2020note}
A.~Nahum, Note on wess-zumino-witten models and quasiuniversality in 2+ 1
  dimensions, \emph{Physical Review B}. {\bf 102}\penalty0 (20), \penalty0
  201116,  (2020).

\bibitem{zhao2020multicritical}
B.~Zhao, J.~Takahashi, and A.~W. Sandvik, Multicritical deconfined quantum
  criticality and lifshitz point of a helical valence-bond phase,
  \emph{Physical Review Letters}. {\bf 125}\penalty0 (25), \penalty0 257204,
  (2020).

\bibitem{liu2022emergence}
W.-Y. Liu, J.~Hasik, S.-S. Gong, D.~Poilblanc, W.-Q. Chen, and Z.-C. Gu,
  Emergence of gapless quantum spin liquid from deconfined quantum critical
  point, \emph{Physical Review X}. {\bf 12}\penalty0 (3), \penalty0 031039,
  (2022).

\bibitem{lu2021self}
D.-C. Lu, C.~Xu, and Y.-Z. You, Self-duality protected multicriticality in
  deconfined quantum phase transitions, \emph{Physical Review B}. {\bf
  104}\penalty0 (20), \penalty0 205142,  (2021).

\bibitem{yang2022quantum}
J.~Yang, A.~W. Sandvik, and L.~Wang, Quantum criticality and spin liquid phase
  in the shastry-sutherland model, \emph{Physical Review B}. {\bf 105}\penalty0
  (6), \penalty0 L060409,  (2022).

\bibitem{shackleton2021deconfined}
H.~Shackleton, A.~Thomson, and S.~Sachdev, Deconfined criticality and a gapless
  z 2 spin liquid in the square-lattice antiferromagnet, \emph{Physical Review
  B}. {\bf 104}\penalty0 (4), \penalty0 045110,  (2021).

\bibitem{zhu2023uncovering}
W.~Zhu, C.~Han, E.~Huffman, J.~S. Hofmann, and Y.-C. He, Uncovering conformal
  symmetry in the 3d ising transition: State-operator correspondence from a
  quantum fuzzy sphere regularization, \emph{Phys. Rev. X}. {\bf 13}, \penalty0
  021009 (Apr, 2023).
\newblock \doi{10.1103/PhysRevX.13.021009}.
\newblock URL \url{https://link.aps.org/doi/10.1103/PhysRevX.13.021009}.

\bibitem{kane2005a}
C.~L. Kane and E.~J. Mele, Quantum spin hall effect in graphene, \emph{Physical
  Review Letter}. {\bf 95}, \penalty0 226801,  (2005).

\bibitem{kane2005b}
C.~L. Kane and E.~J. Mele, $\mathrm{Z_2}$ topological order and the quantum
  spin hall effect, \emph{Physical Review Letter}. {\bf 95}, \penalty0 146802,
  (2005).

\bibitem{grover2008topological}
T.~Grover and T.~Senthil, Topological spin hall states, charged skyrmions, and
  superconductivity in two dimensions, \emph{Physical review letters}. {\bf
  100}\penalty0 (15), \penalty0 156804,  (2008).

\bibitem{khalaf2021charged}
E.~Khalaf, S.~Chatterjee, N.~Bultinck, M.~P. Zaletel, and A.~Vishwanath,
  Charged skyrmions and topological origin of superconductivity in magic-angle
  graphene, \emph{Science advances}. {\bf 7}\penalty0 (19), \penalty0 eabf5299,
   (2021).

\bibitem{christos2020superconductivity}
M.~Christos, S.~Sachdev, and M.~S. Scheurer, Superconductivity, correlated
  insulators, and wess--zumino--witten terms in twisted bilayer graphene,
  \emph{Proceedings of the National Academy of Sciences}. {\bf 117}\penalty0
  (47), \penalty0 29543--29554,  (2020).

\bibitem{liu2019superconductivity}
Y.~Liu, Z.~Wang, T.~Sato, M.~Hohenadler, C.~Wang, W.~Guo, and F.~F. Assaad,
  Superconductivity from the condensation of topological defects in a quantum
  spin-hall insulator, \emph{Nature Communications}. {\bf 10}\penalty0 (1),
  \penalty0 2658,  (2019).

\bibitem{chatterjee2022skyrmion}
S.~Chatterjee, M.~Ippoliti, and M.~P. Zaletel, Skyrmion superconductivity: Dmrg
  evidence for a topological route to superconductivity, \emph{Physical Review
  B}. {\bf 106}\penalty0 (3), \penalty0 035421,  (2022).

\bibitem{thorngren2021intrinsically}
R.~Thorngren, A.~Vishwanath, and R.~Verresen, Intrinsically gapless topological
  phases, \emph{Physical Review B}. {\bf 104}\penalty0 (7), \penalty0 075132,
  (2021).

\bibitem{ma2022edge}
R.~Ma, L.~Zou, and C.~Wang, Edge physics at the deconfined transition between a
  quantum spin hall insulator and a superconductor, \emph{SciPost Physics}.
  {\bf 12}\penalty0 (6), \penalty0 196,  (2022).

\bibitem{hohenadler2022thermodynamic}
M.~Hohenadler, Y.~Liu, T.~Sato, Z.~Wang, W.~Guo, and F.~F. Assaad,
  Thermodynamic and dynamical signatures of a quantum spin hall insulator to
  superconductor transition, \emph{Physical Review B}. {\bf 106}\penalty0 (2),
  \penalty0 024509,  (2022).

\bibitem{kevinQSH}
K.~{Slagle}, Y.-Z. {You}, and C.~{Xu}, {Exotic quantum phase transitions of
  strongly interacting topological insulators}, \emph{\prb}. 91\penalty0
  (11):\penalty0 115121 (Mar., 2015).
\newblock \doi{10.1103/PhysRevB.91.115121}.

\bibitem{so4qsh}
Y.-Y. {He}, H.-Q. {Wu}, Y.-Z. {You}, C.~{Xu}, Z.~Y. {Meng}, and Z.-Y. {Lu},
  {Bona fide interaction-driven topological phase transition in correlated
  symmetry-protected topological states}, \emph{\prb}. 93\penalty0
  (11):\penalty0 115150 (Mar., 2016).
\newblock \doi{10.1103/PhysRevB.93.115150}.

\bibitem{jiang2019ising}
S.~Jiang and O.~Motrunich, Ising ferromagnet to valence bond solid transition
  in a one-dimensional spin chain: Analogies to deconfined quantum critical
  points, \emph{Physical Review B}. {\bf 99}\penalty0 (7), \penalty0 075103,
  (2019).

\bibitem{roberts2019deconfined}
B.~Roberts, S.~Jiang, and O.~I. Motrunich, Deconfined quantum critical point in
  one dimension, \emph{Physical Review B}. {\bf 99}\penalty0 (16), \penalty0
  165143,  (2019).

\bibitem{huang2019emergent}
R.-Z. Huang, D.-C. Lu, Y.-Z. You, Z.~Y. Meng, and T.~Xiang, Emergent symmetry
  and conserved current at a one-dimensional incarnation of deconfined quantum
  critical point, \emph{Physical Review B}. {\bf 100}\penalty0 (12), \penalty0
  125137,  (2019).

\bibitem{zhang2023exactly}
C.~Zhang and M.~Levin, Exactly solvable model for a deconfined quantum critical
  point in 1d, \emph{Physical Review Letters}. {\bf 130}\penalty0 (2),
  \penalty0 026801,  (2023).

\bibitem{mudry2019quantum}
C.~Mudry, A.~Furusaki, T.~Morimoto, and T.~Hikihara, Quantum phase transitions
  beyond landau-ginzburg theory in one-dimensional space revisited,
  \emph{Physical Review B}. {\bf 99}\penalty0 (20), \penalty0 205153,  (2019).

\bibitem{su2023boundary}
L.~Su, Boundary criticality via gauging finite subgroups: a case study on the
  clock model, \emph{arXiv preprint arXiv:2306.02976}.  (2023).

\bibitem{prakash2022multiversality}
A.~Prakash, M.~Fava, and S.~Parameswaran, Multiversality and unnecessary
  criticality in one dimension, \emph{arXiv preprint arXiv:2209.00037}.
  (2022).

\bibitem{jian2020generic}
C.-M. Jian and C.~Xu, Generic unnecessary quantum critical points with minimal
  degrees of freedom, \emph{Physical Review B}. {\bf 101}\penalty0 (3),
  \penalty0 035118,  (2020).

\bibitem{shu2023equilibration}
Y.-R. Shu, S.-K. Jian, A.~W. Sandvik, and S.~Yin, Equilibration of topological
  defects at the deconfined quantum critical point, \emph{arXiv preprint
  arXiv:2305.04771}.  (2023).

\bibitem{hermele2004stability}
M.~Hermele, T.~Senthil, M.~P. Fisher, P.~A. Lee, N.~Nagaosa, and X.-G. Wen,
  Stability of u (1) spin liquids in two dimensions, \emph{Physical Review B}.
  {\bf 70}\penalty0 (21), \penalty0 214437,  (2004).

\bibitem{zou2021stiefel}
L.~Zou, Y.-C. He, and C.~Wang, Stiefel liquids: Possible non-lagrangian quantum
  criticality from intertwined orders, \emph{Physical Review X}. {\bf
  11}\penalty0 (3), \penalty0 031043,  (2021).

\bibitem{cui2023proximate}
Y.~Cui, L.~Liu, H.~Lin, K.-H. Wu, W.~Hong, X.~Liu, C.~Li, Z.~Hu, N.~Xi, S.~Li,
  et~al., Proximate deconfined quantum critical point in srcu2 (bo3) 2,
  \emph{Science}. {\bf 380}\penalty0 (6647), \penalty0 eadc9487,  (2023).

\bibitem{song2023unconventional}
T.~Song, Y.~Jia, G.~Yu, Y.~Tang, P.~Wang, R.~Singha, X.~Gui, A.~J. Uzan,
  M.~Onyszczak, K.~Watanabe, T.~Taniguchi, R.~J. Cava, L.~M. Schoop, N.~P. Ong,
  and S.~Wu.
\newblock Unconventional superconducting quantum criticality in monolayer wte2,
   (2023).

\bibitem{lee2022landau}
J.~Y. Lee, J.~Ramette, M.~A. Metlitski, V.~Vuletic, W.~W. Ho, and S.~Choi,
  Landau-forbidden quantum criticality in rydberg quantum simulators,
  \emph{arXiv preprint arXiv:2207.08829}.  (2022).

\bibitem{coleman2001topical}
P.~Coleman, C.~P{\'e}pin, Q.~Si, and R.~Ramazashvili, Topical review: How do
  fermi liquids get heavy and die?, \emph{Journal of Physics Condensed Matter}.
  {\bf 13}\penalty0 (35), \penalty0 R723--R738,  (2001).

\bibitem{lohneysen2007fermi}
H.~v. L{\"o}hneysen, A.~Rosch, M.~Vojta, and P.~W{\"o}lfle, Fermi-liquid
  instabilities at magnetic quantum phase transitions, \emph{Reviews of Modern
  Physics}. {\bf 79}\penalty0 (3), \penalty0 1015,  (2007).

\bibitem{si2001locally}
Q.~Si, S.~Rabello, K.~Ingersent, and J.~L. Smith, Locally critical quantum
  phase transitions in strongly correlated metals, \emph{Nature}. {\bf
  413}\penalty0 (6858), \penalty0 804--808,  (2001).

\bibitem{senthil2004weak}
T.~Senthil, M.~Vojta, and S.~Sachdev, Weak magnetism and non-fermi liquids near
  heavy-fermion critical points, \emph{Physical Review B}. {\bf 69}\penalty0
  (3), \penalty0 035111,  (2004).

\end{thebibliography}

\end{document}